\documentclass[twocolumn,aps,floatfix,amsmath,amssymb,pre]{revtex4-1}
\usepackage{graphicx}
\usepackage{bbm}
\usepackage{bm}
\usepackage{enumitem}
\usepackage{longtable}

\newcommand{\abs}[1]{\lvert#1\rvert}
\newcommand{\e}[1]{\langle {#1}\rangle}
\newcommand{\be}{\begin{equation}}
\newcommand{\ee}{\end{equation}}
\newcommand{\beqn}{\begin{eqnarray}}
\newcommand{\eeqn}{\end{eqnarray}}
\newcommand{\ket}[1]{{| #1 \rangle}}
\newcommand{\bra}[1]{{\langle #1 |}}

\begin{document}
\title{Tensor network renormalization group study of spin-1 random Heisenberg chains}

\author{Zheng-Lin Tsai}
\author{Pochung Chen}
\email{pcchen@phys.nthu.edu.tw}
\affiliation{Department of Physics, National Tsing Hua University, Hsinchu 30013, Taiwan}
\author{Yu-Cheng Lin}
\email{yc.lin@nccu.edu.tw}
\affiliation{Graduate Institute of Applied Physics, National Chengchi University, Taipei 11605, Taiwan}

\begin{abstract}
We use a tensor network strong-disorder renormalization
group (tSDRG) method to study spin-1 random Heisenberg antiferromagnetic chains.  
The ground state of the clean spin-1 Heisenberg chain with uniform nearest-neighbor couplings is a
gapped phase known as the Haldane phase. Here we consider disordered chains
with random couplings, in which the Haldane gap closes in the strong disorder
regime. As the randomness strength is increased further and exceeds a certain
threshold, the random chain undergoes a phase transition to a critical
random-singlet phase. The strong-disorder renormalization group method
formulated in terms of a tree tensor network provides an efficient tool for
exploring ground-state properties of disordered quantum many-body systems.
Using this method we detect the quantum critical point between the gapless
Haldane phase and the random-singlet phase via the disorder-averaged string
order parameter. We determine the critical exponents related to the average
string order parameter, the average end-to-end correlation function and the
average bulk spin-spin correlation function, both at the critical point and in
the random-singlet phase. Furthermore, we study energy-length scaling
properties through the distribution of energy gaps for a finite chain.  Our
results are in closer agreement with the theoretical predictions than what was
found in previous numerical studies. As a benchmark, a comparison between tSDRG results
for the average spin correlations of the spin-1/2 random Heisenberg chain with those
obtained by using unbiased zero-temperature QMC method is also provided.
\end{abstract}
\date{\today}

\maketitle
\section{Introduction}
\label{sec:intro}

Impurities of different kinds are often naturally contained in real materials
or are introduced by doping.  The effects of disorder and inhomogeneity present
in materials can alter the low-temperature properties dramatically, especially
near quantum critical points; these effects include destruction of quantum
criticality, divergence of dynamic critical exponent, and quantum Griffiths
singularities. Furthermore, there are a number of novel phases emerging from
the interplay between disorder, interactions and quantum fluctuations;
prominent examples for such phases are the many-body localized phase~\cite{MBL}
and certain types of quantum spin-liquids~\cite{QSL_Rev,Kimchi,PRX,RQSL_Rev}.

Numerical studies of disorder systems are notoriously difficult,
mainly because (i) disorder is often accompanied by a long relaxation time and
rough energy landscape, which leads some standard algorithms having a tendency
to get stuck in local minima; (ii) there is a lack of translational symmetry,
which makes the ”infinite version” of tensor-network based approaches impractical. 
On the other hand, the strong disorder renormalization group
(SDRG) designed specifically for disordered systems provides an analytical tool
to capture asymptotically exact ground-state properties for a number of
one-dimensional (1D) systems~\cite{MDH,Fisher50,Fisher51,SDRG_Rev1,SDRG_Rev2} 
and can also be implemented numerically on more
complex systems, including systems with geometrical frustration, as long as the
disorder is sufficiently strong~\cite{Ladder,2DAF,Bilayer,2DAnyon,1DJQ,1DJ1J2}.

The SDRG method was first introduced for solving the random spin 1/2 Heisenberg
antiferromagnetic chain~\cite{MDH,Fisher50}. The iterative SDRG procedure consists of locking the
strongest coupled pair of spins into a singlet (a valence bond) 
and renormalizing the coupling between the neighboring spins by perturbation
theory.
Repeating these steps ultimately leads to an approximate ground state---the random-singlet (RS) state~\cite{Fisher50},
in which each spin is paired into a singlet with another spin which may be arbitrarily far away.
Those long-ranged singlets formed by widely separated spins are rare; however, they dominate the {\it average} 
spin-spin correlations that decay asymptotically with distance $L$ as an inverse-square form $L^{-2}$.
By contrast, a typical pair of spins is not in the same singlet and 
has only weak correlations that fall off exponentially with the square root of their distance.
The energy-length scaling can be obtained by considering the energy scale 
(i.e. the strength of the renormalized coupling) of a singlet with length $L$, 
yielding 
\be
  -\ln \epsilon \sim L^\psi\,.
  \label{eq:E_L_sc}
\ee 
with $\psi=1/2$.
This type of scaling, 
which is very different from the standard scaling $\epsilon\sim L^{-z}$,
implies that the dynamical exponent diverges: $z=\infty$. With the diverging dynamical exponent the RS fixed point 
is a so-called infinite-randomness fixed point and it is a stable fixed point for the spin-1/2 chain
with arbitrarily weak randomness.

Unlike the application for the spin-1/2 chain, the conventional SDRG scheme
breaks down for Heisenberg chains with higher spins $S > 1/2$ in the regime of weak randomness.
This is because renormalized couplings for $S > 1/2$ may become stronger than the decimated couplings during RG, 
which makes perturbation theory invalid~\cite{Boechat}. 
Nevertheless, the SDRG method is applicable to $S > 1/2$ chains in the limit of strong disorder,
where the systems are in the RS phase too.
More generally, a higher-$S$ random chain can be mapped to an effective $S=1/2$ chain and
can then be treated by extended SDRG approaches even for weaker randomness. 
In previous SDRG studies on effective $S=1/2$ models~\cite{Hyman,Monthus,Damle,Damle_Huse,S23}, 
second order phase transitions
from weak randomness phases to the spin-$S$ RS phase were found;
the critical points are infinite randomness fixed points 
that are not in the same RS universality class~\cite{Damle,Damle_Huse,S23}.

In this paper we will use a tree tensor network algorithm in combination with
the idea of the SDRG to examine the ground state properties of the $S=1$ random Heisenberg antiferromagnetic chain.   
The ground state of the $S=1$ chain in the absence of randomness is
in the so-called Haldane phase~\cite{Haldane}, which is a gapped phase and
possesses string topological order~\cite{Nijs}. 
The Haldane phase and its topological order are
stable against weak randomness~\cite{Hyman}.
Here we will focus on the ground-state phases where the energy gap is destroyed by randomness;
they are gapless Haldane phase with hidden topological order, the spin-1 RS phase and 
the critical point between these two phases.

\section{The model}
\label{sec:model}

\begin{figure}
\centering
\resizebox{0.9\columnwidth}{!}{\includegraphics{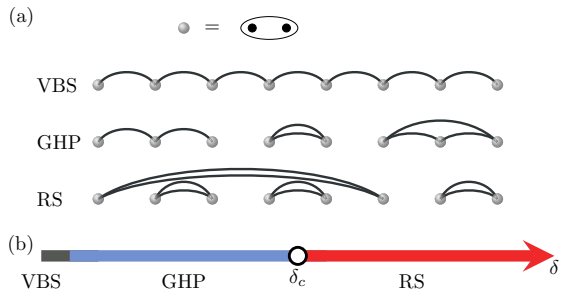}}
\caption{
(a) Possible ground states of the spin-1 random Heisenberg chain. 
The spin-1 on each site (represented by a gray shaded sphere)
is composed of the symmetric combination of two $S=1/2$ spins (solid circles).
Without disorder or at very low disorder, the ground state is a gapped valence-bond solid (VBS),
in which each site forms a singlet-1/2 to its right and to its left.
With sufficiently strong disorder, the gap is destroyed due to defects in the VBS structure. 
This gapless Haldane phase (GHP) is a Griffiths phase with topological order. 
At strong disorder, the ground state undergoes a phase transition at $\delta_c$ and
becomes a critical
random singlet (RS) phase, where each $S=1$ spin forms a singlet pair with another
$S=1$ spin, which may span arbitrarily long distances. 
(b) Phase diagram depending on the randomness strength $\delta$.
}
\label{fig:phases}       
\end{figure}

We study the spin-1 random antiferromagnetic Heisenberg chain, described by the Hamiltonian:
\begin{equation}
        \label{eq:H1}
        H =  \sum_{i} J_{i}\vec{S}_{i}\cdot\vec{S}_{i+1} ,
\end{equation}
where $\vec{S}_i$ is the spin-1 operator at site $i$
and $J_{i} > 0$ is a random antiferromagnetic coupling. 
We use the following distribution of the random couplings:
\begin{equation}
        \label{eq:J_dis}
        \pi_{\delta}(J) = \frac{1}{\delta}J^{-1 + \frac{1}{\delta}} \quad \textrm{for} \quad 0 < J \leq 1,
\end{equation}
where $\delta$, being the standard deviation of $\ln(J)$, parameterizes the strength of disorder.
This power-law distribution corresponds to a uniform distribution when $\delta=1$ 
while it becomes highly singular at the origin for $\delta\gg 1$.

The Haldane ground state in the absence of randomness (i.e. $\delta=0$) is well
described by the valence-bond solid (VBS) state~\cite{AKLT}, in which each spin-1
is considered to be a symmetric combination of two spins-1/2 and a singlet (a
valence bond) is formed between two spin-1/2 objects on neighboring sites
(see Figure~\ref{fig:phases}(a)). Such a VBS state has a
long-range topological order characterized by hidden staggered $S^z=+1,\,-1$
configuration after removing all sites with $S^z=0$.
This hidden topological order
can be probed by the string order parameter~\cite{Nijs}
\begin{equation}
        O^{z}_{j,k} = -\left \langle S_{j}^{z} \exp(i \pi \sum_{l = j + 1}^{k - 1} S_{l}^{z} ) S_{k}^{z} \right\rangle \,,
        \label{eq:Oz}
\end{equation}
where $S^z_j$ is the $z$ component of the spin operator at site $j$.

In the presence of randomness, the spin-1 chain exhibits various ground-state
phases, depending on the strength of randomness.  
A schematic phase diagram,
based on previous studies~\cite{Hyman,Monthus,Damle}, is shown in
Fig.~\ref{fig:phases}(b).  
With sufficiently strong randomness (for any
$\delta>0$ using the power-law distribution in
Eq.~(\ref{eq:J_dis})~\cite{DMRG1} ), the gap vanishes due to
defects occurring in the VBS structure (see Figure ~\ref{fig:phases}(a)) but
the topological order can survive up to a critical value $\delta_c$. This
gapless Haldane phase is a Griffiths phase with short-range spatial
correlations and a power-law density of states for low-energy
excitations~\cite{Damle}: $\rho(\epsilon)\sim \epsilon^{-1+1/z}$.
The dynamical exponent $z$, which appears in $\rho( \epsilon)$, varies
continuously in the Griffiths phase with the distance from the location of
$\delta_c$. The power-law density of states $\rho( \epsilon)$ results in
power-law singularities in some thermodynamical quantities, such as the local
susceptibility $\chi_\text{local}$ which behaves as $\chi_\text{local}\sim
T^{-1+1/z}$ at low temperature and diverges at $T=0$ if $z>1$. 

For $\delta>\delta_c$, the system enters a critical spin-1 RS phase, where
singlets connect spins-1 over arbitrarily long distances and the string
topological order vanishes.  This spin-1 RS phase is analogous to
the spin-1/2 RS phase. 
Some results obtained for the spin-1/2 RS phase are valid for the spin-1
case too. For example, here the length-energy scaling 
obeys the form of Eq.~(\ref{eq:E_L_sc})  with $\psi=1/2$.
The spin correlation function, defined as 
\be 
   C(r)=(-1)^r\e{\vec{S}_i\cdot \vec{S}_{i+r}}\,,
   \label{eq:C_def}
\ee
typically behaves as
\be
  -\ln C(r)\sim r^\psi\,,
  \label{eq:C_typ}
\ee
while the average spin
correlations decay asymptotically with distance $r$:
\be
   \overline{C}(r)\sim \frac{1}{r^2}\,,
\ee
where the overline denotes averaging over the randomness.
Also the average end-to-end correlation function in an open chain of length $L$ 
decays algebraically as~\cite{Fisher_Young}
\be
    \overline{C}_{1}(L) \sim \frac{1}{L}\,,
   \label{eq:C_1}
\ee
although typical end-to-end correlations are exponentially weak and
broadly distributed. The distinction between average and typical
values is in fact one of the main features of an infinite-randomness fixed point.

The critical point at $\delta=\delta_c$, separating the gapless Haldane phase and the RS phase, 
has turned out to be an infinite-randomness fixed point, too. 
It has a similar energy-length scaling relation in the form of Eq.~(\ref{eq:E_L_sc});
however, the associated exponent here is $\psi=\psi_c=1/3$. 
This critical point is a multicritical point at which three topologically distinct 
phases meet; these phases are classified by the numbers of valence bonds formed
across the even and odd links of the lattice~\cite{Damle,Damle_Huse}.

As the disorder strength approaches the critical point from the gapless Haldane phase,
the average string order parameter, given by 
\be
  O^z(r=\abs{j-k})=\overline{O}^z_{j,k}\,,
\ee 
decays to zero as
\begin{equation}
\label{eq:Oz_d}
        O^{z} \sim (\delta_{c} - \delta)^{2\beta}\,,
\end{equation} 
with a universal exponent $\beta$ given by~\cite{Monthus}
\be
    \beta=\frac{2(3-\sqrt{5})}{\sqrt{13}-1}\approx0.5864\,.
    \label{eq:beta} 
\ee 
At the same time, the average correlation length grows continuously from 
a finite value to infinite when approaching criticality:
\begin{equation}
       \xi\sim (\delta_{c} - \delta)^{-\nu}\,,
       \label{eq:xi}
\end{equation}
with~\cite{Hyman}
\be
   \nu=\frac{6}{\sqrt{13}-1}\approx 2.3028\,.
   \label{eq:nu}
\ee
Therefore, at the critical point $O^{z}$ decays algebraically with the length (the distance) as
\begin{equation}
\label{eq:Oz2}
        O^{z}(r) \sim r^{-\eta_{\text{st}}}\,,
\end{equation}
where the critical exponent $\eta_{\text{st}}$ is related to $\beta$ via
\begin{equation}
        \label{eq:beta_eta_nu}
        \eta_{\text{st}}=2\beta/\nu\approx 0.5093\,.
\end{equation}
Since the critical point is not in the random-singlet universality class,
the critical exponents $\eta,\,\eta_1$ for the power-law decaying average spin correlations 
($\overline{C}(r)\sim 1/r^\eta$)
and average end-to-end correlations ($\overline{C}_1\sim 1/r^{\eta_1}$) are not expected
to be the same with those at the RS fixed point. 
There have been so far no theoretical conjectures  
about the exponents $\eta$ and $\eta_1$ at this critical point.

Considerable numerical efforts using the density matrix renormalization~\cite{DMRG1,Hida,DMRG2} and 
quantum Monte Carlo simulations~\cite{QMC} have been devoted to examine the
theoretical predictions and gain more insights into universal features of the
spin-1 random chain. However, there remain discrepancies between some numerical
results. In this paper we use a tree tensor-network algorithm in combination
with the SDRG scheme to re-study the spin-1 random chain. In the following section 
we describe the scheme of this tensor network strong-disorder
renormalization group (tSDRG) method~\cite{tSDRG0,tSDRG1,tSDRG2}.

\section{Tensor network strong-disorder renormalization group}
\label{sec:method}

The tSDRG method is, in essence, a renormalization of the Hamiltonian written
as Matrix Product Operators (MPOs)~\cite{tSDRG1,tSDRG2}. 
In the computation of quantum many-body
systems, matrix product representation is a powerful tool to reduce execution
time and memory usage via the decomposition of a big tensor, which represents a state or
an operator, into a set of small local tensors~\cite{Scholl,Verstraete,Orus1,Orus2}.

For a spin chain of length $L$ with open boundary conditions (OBC), the
Hamiltonian can be decomposed into a matrix product form written as
\be
     {H}=W^{[1]}W^{[2]}\cdots W^{[L]}\,,
\ee
with
\be
    W^{[i]}=\sum_{\sigma_i,\sigma'_i} W^{\sigma_i,\sigma'_i}\ket{\sigma_i}\bra{\sigma'_i}\,,
\ee
where $\sigma_i$ labels the spin state at site $i$. 
To construct the MPO we rewrite the Hamiltonian of the Heisenberg chain in terms of the ladder operator 
$S^{\pm}=S^x\pm i S^y$:
\begin{equation}
       \label{eq:H2}
       {H} =  \sum_{i} J_{i}[\frac{1}{2}({S}_{i}^{+} {S}_{i+1}^{-} + {S}_{i}^{-} {S}_{i+1}^{+} ) + {S}_{i}^{z} {S}_{i+1}^{z}]\,.
\end{equation}
This Hamiltonian has the following $W^{[i]}$-tensors for sites in the bulk,

\begin{equation}
\label{eq:MPO2}
W^{[i]} =
\begin{pmatrix}
        \mathbbm{1} & 0 & 0 & 0 & 0 \\
        {S}_{i}^{+} & 0 & 0 & 0 & 0 \\
        {S}_{i}^{-} & 0 & 0 & 0 & 0 \\
        {S}_{i}^{z} & 0 & 0 & 0 & 0 \\
        0 & (J_{i}/2){S}_{i}^{-} & (J_{i}/2){S}_{i}^{+} & J_{i}{S}_{i}^{z} & \mathbbm{1} \\
\end{pmatrix}\ ,
\end{equation}
and for the edge sites,
\begin{equation}
\label{eq:MPO1}
W^{[1]} =
\begin{pmatrix}
        0\; & (J_{1}/2){S}_{1}^{-}\; & (J_{1}/2){S}_{1}^{+}\; & J_{1}{S}_{1}^{z}\; & \mathbbm{1} \\
\end{pmatrix}\ ,
\end{equation}

\begin{equation}
\label{eq:MPO3}
W^{[L]} =
\begin{pmatrix}
        \mathbbm{1}  \\
        {S}_{L}^{+}  \\
        {S}_{L}^{-} \\
        {S}_{L}^{z} \\
        0  \\
\end{pmatrix}\ .
\end{equation}

For a chain with periodic boundary conditions (PBC), the MPO tensors for $i=1\cdots
L$ are all bulk tensors as given in Eq.~(\ref{eq:MPO2}), where the coupling
$J_L$ links between two end sites $L$ and $1$.  

The tSDRG procedure consists of iteratively locating a
local Hamiltonian with the largest energy gap, and truncating its Hilbert space
to the subspace spanned by the eigenvectors corresponding to the eigenvalues
below the gap.  
A local Hamiltonian in the original Hamiltonian 
takes the form of a two-site Hamiltonian:
\begin{equation}
        {h}^{(i, i+1)}=J_i \vec{S}_{i}\cdot\vec{S}_{i+1}\,,
\end{equation}
and is encoded in the matrix element of $W^{[i]} W^{[i+1]}$ as follows: 
\begin{equation}
\label{eq:merge2}
W^{[i]} W^{[i+1]} =
\begin{pmatrix}
        \mathbbm{1}  & 0 & 0 & 0 & 0 \\
        {S}_{i}^{+}  & 0 & 0 & 0 & 0 \\
        {S}_{i}^{-}  & 0 & 0 & 0 & 0 \\
        {S}_{i}^{z}  & 0 & 0 & 0 & 0 \\
        h^{(i,i+1)} & \frac{J_{i+1}}{2}{S}_{i+1}^{-} & \frac{J_{i+1}}{2} {S}_{i+1}^{+} &  J_{i+1}{S}_{i+1}^{z} & \mathbbm{1} \\
\end{pmatrix}\,.
\end{equation}
In each RG iteration, we compute the energy spectrum of each local Hamiltonian and identify the energy gap
$\Delta\epsilon^{(i, i+1)}$, which is measured as the difference between the highest energy of the 
$\chi$-lowest energy states that would be kept and the higher multiplets that would be discarded.  
We then choose the local Hamiltonian with the largest energy gap and
coarse-grain the tensors on the two sites into a new single-site MPO tensor
using the $\chi$-lowest energy states. 
The process is iterated until the whole system is coarse-grained into a single site.
The full RG process is summarized as follows~\cite{tSDRG1}:
\begin{description}
\item[{\bf (i)}] Decompose the Hamiltonian into MPO blocks; 
each block contains one site. 
\item[{\bf (ii)}] Find the largest energy gap of the local Hamiltonian $h^{(i,i+1)}$ for each pair of nearest-neighbor blocks; 
here the gap is $\Delta \epsilon^{(i,i+1)}=2J_i$ for each $h^{(i,i+1)}$ in the original Hamiltonian with $S=1$.
\item[{\bf (iii)}] Find the the pair of blocks with the largest gap $\Delta \epsilon^{(m,m+1)}$ and
contract the MPO tensors $W^{[m]}$ and $W^{[m+1]}$  for these blocks. 
\item[{\bf (iv)}] Identify 
$\chi' (\le \chi)$ lowest energy states $\ket{\Psi_1},\ket{\Psi_2},$ $\cdots,\ket{\Psi_{\chi'}}$ that will be kept; 
here the {\it bond dimension} $\chi$ is an input parameter setting the upper bound of the number of states to be kept, 
and the actual number $\chi'$ is adjusted such that the kept states form full SU(2) multiplets. 
\item[{\bf (v)}] Build a three-leg isometric tensor $V$ using the $\chi'$ lowest energy eigenstates:
\be
   V=
    \begin{pmatrix}
     \ket{\Psi_1} & \ket{\Psi_2} & \cdots & \ket{\Psi_{\chi'}} 
    \end{pmatrix}\,,  
\ee
which satisfies that $V^\dagger V=\mathbbm{1} \neq VV^{\dagger}$.
\item[{\bf (vi)}]
Renormalize the pair of blocks with the largest gap by contracting the
two-block tensors with $V$ and $V^\dagger$:
\be
      \widetilde{W}^{[m]}=V^\dagger W^{[m]}  W^{[m+1]} V\,.
\ee
\item[{\bf (vii)}] Repeat steps {\bf (iii)} to {\bf (vi)} until there remains one single MPO tensor.
In the final MPO there is only one matrix element which represents the final local Hamiltonian $h_f$.
\end{description}

\begin{figure}
\centering
\resizebox{0.8\columnwidth}{!}{\includegraphics{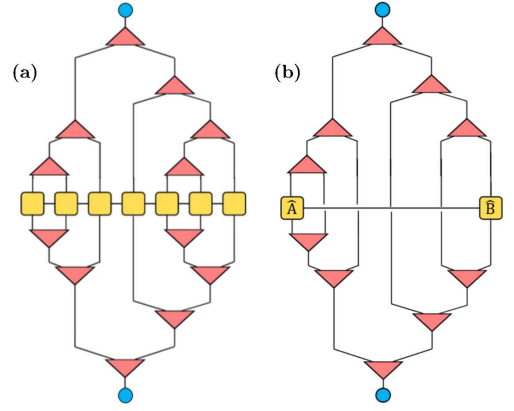}}
\caption{
(a) The tSDRG algorithm can be seen as a binary tree tensor network with an inhomogeneous structure.
The system is partitioned into blocks using MPO formalism; the yellow boxes represent the W-tensors,
the vertical lines are physical indices, and the horizontal lines represent the bond indices.
The triangles denote isometric tensors and the circles represent the ground-state eigenvector of the final block resulting from the RG procedure.
The RG iteration proceeds upwards in the vertical dimension.
The part below the W-tensors is the conjugate of the upper part;
(b) Part of the tensor network in (a) used to calculate the ground-state expectation value of $\hat{A}\hat{B}$;
since $V^\dagger V=\mathbbm{1}$ only those isometric tensors linking to the operators $\hat{A}$
and $\hat{B}$ are considered.}
\label{fig:tSDRG}       
\end{figure}

The full tSDRG algorithm can be seen as an inhomogeneous binary tree tensor network, 
composed of isometric tensors that each merges two blocks into an effective block, 
and one top tensor representing the final remaining block 
((see Fig.~\ref{fig:tSDRG}(a)).   
The ground-state expectation value of some observable can be obtained by contracting the operator
with isometric tensors and their conjugates until the top tensor (see Fig.~\ref{fig:tSDRG}(b)).

\begin{figure}
\centering
\resizebox{0.8\columnwidth}{!}{\includegraphics{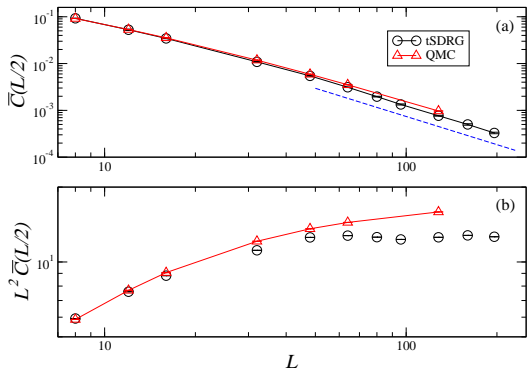}}
\caption{
(a) tSDRG and QMC results for the average spin correlations of the spin-1/2 random chain 
at the largest distance. 
The QMC results are taken from Ref.~\cite{1DJQ}.
The blue dashed line indicates the inverse-square function $1/L^2$.
(b) The correlation functions in (a) multiplied by $L^2$. The result shows the presence 
of a multiplicative logarithmic (log) correction to the $1/L^2$ scaling for the QMC results, 
which increases with distance; the correction in the SDRG results increases
for small $L$ but converges to a constant at large $L$.
}
\label{fig:S05_qmc}       
\end{figure}

Before presenting our results obtained by tSDRG for the spin-1 random chain in
the next section, here, as a test, we first compare the tSDRG result for the average
spin correlations of the spin-1/2 random chain with the nonperturbative QMC result~\cite{1DJQ}.
For this comparison, the random couplings were chosen to be uniformly distributed within 
the range $(0,1]$, corresponding to
$\delta=1$ in Eq.~(\ref{eq:J_dis}) and the disorder-averaged spin correlations at $r=L/2$ in chains with PBC are considered.
In Fig.~\ref{fig:S05_qmc}, the data of the tSDRG with $\chi=30$ (which has achieved convergence, see \ref{app}) are
in good agreement with the QMC data and follow the expected $L^{-2}$ decay for a spin-1/2 RS phase.
For large $L$, a clear deviation from $L^{-2}$ can been seen in the QMC results, which indicates a 
multiplicative logarithmic correction as shown in Fig.~\ref{fig:S05_qmc}(b)
and discussed in Ref.~\cite{1DJQ}.
The tSDRG method, like the conventional SDRG, does not capture the log correction
seen in the QMC calculation. 
Nevertheless, the tSDRG data largely agree with the QMC results and
the tSDRG technique seems to be a promising calculational route to rich ground-state phases
of more complex random spin models, such as higher-$S$ random chains.

\section{Numerical results}
\label{sec:results}

We have used the
Uni10 library~\cite{Uni10} to perform tSDRG calculations. 
In this section we present our tSDRG results for some ground-state observables of the random Heisenberg $S=1$ chain.
These observables include the string order parameter, distributions of energy gaps, spin correlations.
In the following, we discuss the results for these observables, separately.
\subsection{String order parameter}
The string order parameter can be used to identify the critical point between
the gapless Haldane phase where hidden topological order presents and the
RS phase where the hidden order is completely destroyed by strong
disorder.
In our numerical work we calculated the average string order parameter $O^z(r)$,
defined in Eq.~(\ref{eq:Oz}) and Eq.~(\ref{eq:Oz_d}), at the largest distance $r=L/2$
in a closed chain with PBC, for system sizes up to $L=256$ and for various values of $\delta$; 
in each case at least 1000 samples (disorder realizations) were considered and, in addition,
$L/2$ different reference locations in the closed chain were sampled for the
disorder average. The largest bond dimension was $\chi=30$.

\begin{figure}
\centering
\resizebox{0.8\columnwidth}{!}{\includegraphics{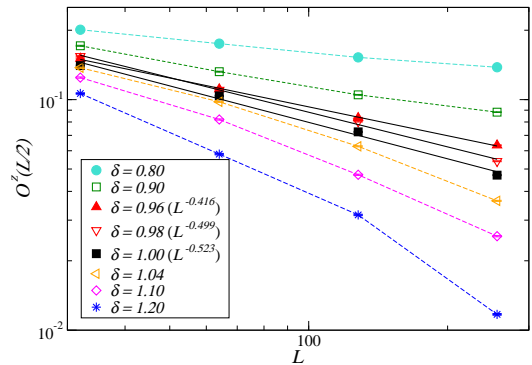}}
\caption{
Average string order parameter $O^z$ at the longest distance $L/2$ for several different chain lengths $L$
with different disorder parameters $\delta$.
Black solid lines on the data for $\delta=0.96, 0.98$ and $1.00$ correspond to a fitting form $O^z=A L^{-\eta_\text{st}}$.
}
\label{fig:O_L}       
\end{figure}

First, in Fig.~\ref{fig:O_L} we show the average string order parameters as
functions of $L$ for various disorder parameters $\delta$ near the transition
point. From the decay behavior of the curves in the log-log plot and a
comparison of the exponent $\eta_{\text{st}}$ with the theoretical conjecture
($\eta_{\text{st}} \approx 0.5093$), it seems reasonable to fix $\delta_c=1$ for our
results, which is also consistent with previous numerical results obtained by
the density matrix renormalization group
(DMRG)~\cite{DMRG1,DMRG2}.  At $\delta_c=1$ we obtain
$\eta_{\text{st}} \approx 0.52$, slightly larger than the theoretical value.

\begin{figure}
\centering
\resizebox{0.8\columnwidth}{!}{\includegraphics{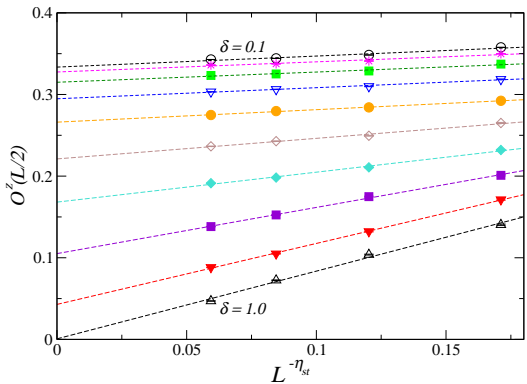}}
\caption{Extrapolations to the thermodynamic limit of the string order parameters for 
disorder strength $\delta=0.1, 0.2, \cdots, 0.9, 1.0$ using
the fitting function in Eq. (\ref{eq:extrapolation_Oz}) with $\eta_\text{st}=0.5093$. The extrapolated value $O^z(\infty)$
for the curve with $\delta=1.0$ approaches zero, indicating that the critical point is located at $\delta_c=1$.
}
\label{fig:O_extra}       
\end{figure}

\begin{figure}
\centering
\resizebox{0.85\columnwidth}{!}{\includegraphics{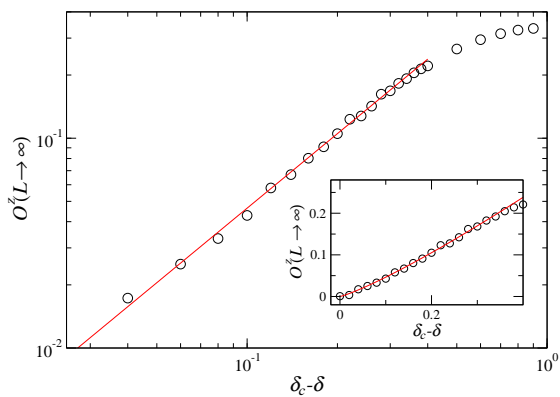}}
\caption{The limiting values $O^z(\infty)$, obtained in Fig.~\ref{fig:O_extra} using Eq.~(\ref{eq:extrapolation_Oz}),  
versus $(\delta_c-\delta)$
with $\delta_c=1$ in a log-log plot. The red dashed line is the best fit using
Eq.~(\ref{eq:Oz_d}) and has an exponent of $2\beta = 1.18$.
The inset shows the data and the fitting function in a linear-linear plot, 
including two points at $\delta_c-\delta=0$ and
$\delta_c-\delta=0.02$ which are not shown in the log-log plot;
the fit deviates from the data noticeably only in the region $\delta_c-\delta\gtrsim 0.38$.
}
\label{fig:O_d}       
\end{figure}

In order to obtain the string order parameter in the thermodynamic limit we
extrapolate the data in the range $\delta\le 1$ for finite sizes to
$L\to\infty$ using the fitting function
\begin{equation}
   \label{eq:extrapolation_Oz}
   O^{z}(r = {L}/{2}) = O^{z}(\infty) + A_L L^{-\eta_{\text{st}}}\,,
\end{equation}
with the theoretical value $\eta_{\text{st}}=0.5093$, as shown in Fig.~\ref{fig:O_extra}.
We have also determined the exponent
$\beta$, defined in Eq.~(\ref{eq:Oz_d}), from $O^{z}(\infty)$ as a function
of $(\delta_c-\delta)$ for $\delta\le \delta_c\, (=1)$ in a log-log plot shown
in Fig.~\ref{fig:O_d} and obtain $2\beta\approx 1.18$, i.e.
$\beta\approx 0.59$.

Finally, we obtain the correlation length exponent
$\nu \approx 2.27$ via Eq.~(\ref{eq:beta_eta_nu})
with our estimated $2\beta=1.18$ and $\eta_{\text{st}}=0.52$.
Our tSDRG results alongside the results from previous DMRG studies in Ref.~\cite{DMRG1}
and Ref.~\cite{DMRG2} are listed in Table \ref{Tab:exponent}.

\begin{table*}[t]
\centering
    \caption{\label{Tab:exponent} Critical exponents for the critical point (CP) and the random-singlet (RS) phase. The
exponents obtained by the simple SDRG are analytical results based on (effective) $S=1/2$ models~\cite{Fisher50,Hyman,Monthus,Fisher_Young}.}
    \begin{tabular}{|p{3cm}| p{1cm} | p{1cm} | p{1cm} | p{1cm} | p{1cm} || p{1cm} | p{1cm} |}
        \hline
			  & \multicolumn{5}{c ||}{ CP ($\delta=\delta_c$) } & \multicolumn{2}{c |}{ RS ($\delta>\delta_c$)}  \\ \hline	
                          & $\eta_\text{st}$ & $\beta$    & $\nu$    & $\eta_1$   &  $\eta$ & $\eta_1$   &  $\eta$   \\ \hline
        Simple SDRG            & 0.5093           & 0.5864     & 2.3028  &   -     &  -     &  1    &   2     \\ \hline
        tSDRG (this work) & 0.52(3)          & 0.59(2)    & 2.3(1)  &  0.70(2)  &   1.62(5)  &  1.1(2)  &   2.03(8) \\ \hline
        DMRG(2005)~\cite{DMRG1} & 0.39(3)    & -          & -       &  0.69(5)  & -     &  0.86(6)   &  -    \\ \hline
        DMRG(2018)~\cite{DMRG2} & 0.21(4)   & 0.24(5)     & 2.3(4)  & -        & -   &  -          &   -     \\ \hline 
    \end{tabular}
\end{table*}

\subsection{Energy gaps}
In this subsection we focus on the distribution of energy gaps. 
From the scaling behavior of the distribution
we can distinguish between a Griffiths phase and an infinite-randomness phase.
In a Griffiths phase the low-lying gaps follow a power-law distribution with an
exponent that is determined by a nonuniversal dynamical exponent, which is randomness dependent.
In an infinite-randomness phase, the dynamical exponent diverges $z\to\infty$ and the
energy gaps are characterized by an extremely broad distribution which becomes
broader with increasing size, even on a logarithmic scale.

\begin{figure}
\centering
\resizebox{0.8\columnwidth}{!}{\includegraphics{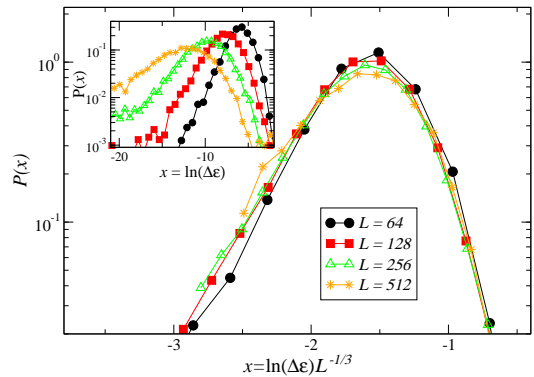}}
\caption{Scaling plot of the distribution of energy gaps at
the critical point ($\delta=1$), assuming $-\ln(\Delta\epsilon)\sim L^\psi$ with $\psi=1/3$.
The inset shows the unscaled distribution of $\ln(\Delta \epsilon)$,
obtained from the lowest-lying energy gap of the renormalized Hamiltonian in the top tensor for 10000 samples for each size.}
\label{fig:gaps_d10}       
\end{figure}
\begin{figure}
\centering
\resizebox{0.8\columnwidth}{!}{\includegraphics{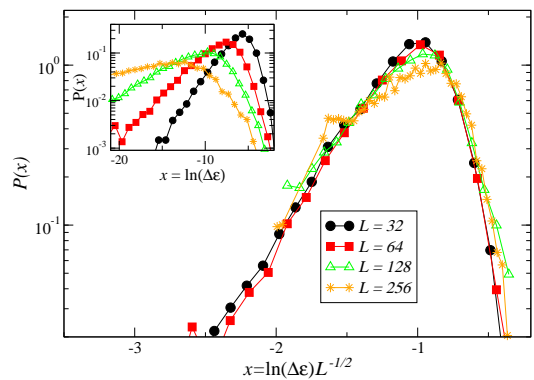}}
\caption{Scaling plot of the distribution of energy gaps in the RS phase ($\delta=1.5$), 
assuming $-\ln(\Delta\epsilon)\sim L^\psi$ with $\psi=1/2$.
The inset shows the unscaled distribution of $\ln(\Delta \epsilon)$,
obtained from the lowest-lying energy gap of the renormalized Hamiltonian in the top tensor for 10000 samples for each size.
}
\label{fig:gaps_d15}       
\end{figure}

We have determined the energy gap, $\Delta\epsilon$, of a sample from the lowest-lying excitation of 
the renormalized Hamiltonian in the top tensor. 
First we examine the distribution of the energy gaps at the critical point, $\delta_c=1.0$, 
and show a scaling plot of the distribution in Fig.~\ref{fig:gaps_d10}.
The distribution which is broadened with increasing $L$,
as shown in the inset of the figure, 
clearly signals an infinite randomness critical point;
the data collapse is achieved by using the scaled variable
\be 
x=-\ln \Delta\epsilon/L^{\psi}
\label{eq:gap_sc}
\ee 
with $\psi=1/3$, in agreement with the theoretical prediction~\cite{Damle_Huse}.
We have also calculated energy gaps for $\delta>\delta_c$.
An example for $\delta=1.5$ is shown in Fig.~\ref{fig:gaps_d15};
here the broad distributions of the logarithmic energy gaps can be
rescaled using the same form in Eq.~(\ref{eq:gap_sc}), but with $\psi=1/2$, 
to achieve the data collapse.  
Here we comment on the poor data collapse of the distributions 
around the maximum for the largest system sizes. 
These numerical errors are caused by the extremely small energy gaps for large $L$ with large $\delta$ 
which make the eigensolver fail to converge;
similar problems were observed in previous numerical works~\cite{tSDRG2,Quench,Zhao,NJP} dealing with
infinite-randomness fixed points.
A possible route to circumventing this numerical instability 
is to use multiple-precision arithmetic, as discussed in Ref.~\cite{NJP,Hoyos}.  

\begin{figure}
\centering
\resizebox{0.8\columnwidth}{!}{\includegraphics{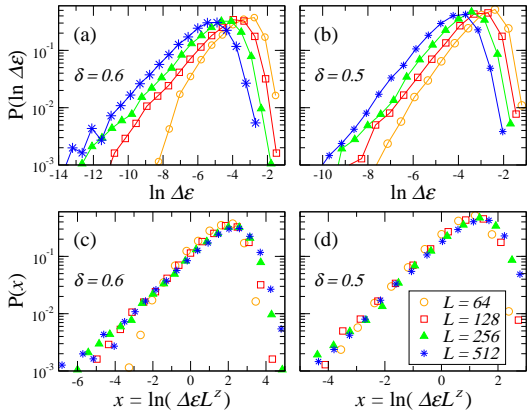}}
\caption{The distribution of energy gaps in the gapless Haldane phase at $\delta=0.6$ (a) and $\delta=0.5$,
collected from 10000 samples for each size.
(c) and (d) are scaling plots of the data in (a) and (b), respectively. Here the dynamical exponent
$z$ is finite. The fit has $z=1.2$ for $\delta=0.6$ and $z=0.87$ for $\delta=0.5$.}
\label{fig:griffiths}       
\end{figure}

With weaker disorder $\delta<1$, 
the width of the gap distribution becomes saturated for $L\to\infty$.
Fig.~\ref{fig:griffiths}(a) and Fig.~\ref{fig:griffiths}(b) show results for $\delta=0.6$
and $\delta=0.5$, respectively;
here the tails of the small gaps for large $L$ tend to a power-law form
consistent with the presence of a Griffiths phase.
The power of the low-energy tail of $P(\ln(\Delta\epsilon))$ is given by
$1/z$~\cite{Rieger_Young,Frechet}. 
From the slope of the power-law tails in Fig.~\ref{fig:griffiths}, we obtain $z=1.2$ and $z=0.87$
for $\delta=0.6$ and $\delta=0.5$, respectively.
Scaling plots using the scaling variable $\Delta\epsilon L^z$,
are shown in Fig.~\ref{fig:griffiths}(c),(d).
The dynamical exponent $z<1$ for $\delta=0.5$ does not lead to divergence
of the local susceptibility (see Sec.~2). 
Therefore, the region where $z<1$, such as $\delta \lesssim 0.5$,
corresponds to the nonsingular region in the gapless Haldane phase, as discussed
in Ref.~\cite{DMRG1}.

\subsection{Spin correlations}
\begin{figure}
\centering
\resizebox{0.8\columnwidth}{!}{\includegraphics{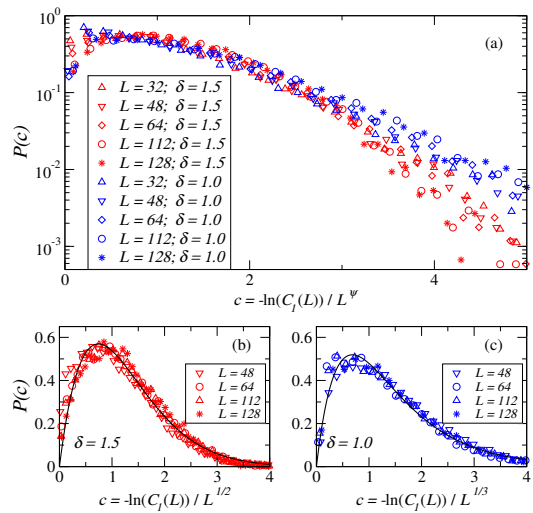}}
\caption{Scaling plots of the distribution of the end-to-end correlation
function at the critical point (blue data) and in the RS phase (red data). 
The correlation functions are rescaled as $-\ln C_1/L^\psi$ with $\psi=1/3$ for  the critical point
and with $1/2$ for the RS phase to achieve data collapse.
The solid lines in the linear-linear plots in (b) and (c) 
are attempts to fit the data-collapsed distributions
to the form in  Eq.~(\ref{eq:P_c}).
}
\label{fig:C1_P}       
\end{figure}
\begin{figure}
\centering
\resizebox{0.8\columnwidth}{!}{\includegraphics{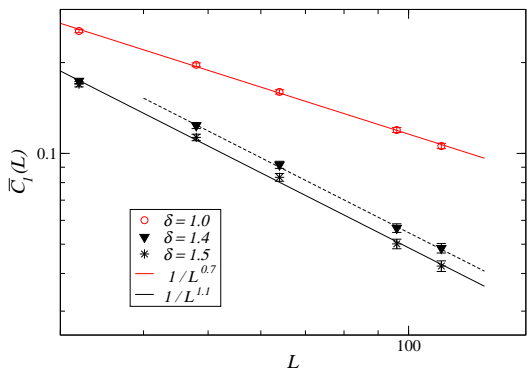}}
\caption{The average of the end-to-end correlation
functions at the critical point ($\delta=1$) and
in the RS phase ($\delta>1$). 
The red line is a fit to the data at $\delta=1.0$.
The black solid line on the data for $\delta=1.5$ is the best fit to the data points from $L=32$ to $L=112$,
which is of the form $a/L^{1.12}$; the black dashed line on the data for $\delta=1.4$
has the same form $\propto 1/L^{1.12}$ but with a different amplitude, which fits well to the data points
in the regime of large $L$ ($L\ge 48$).}
\label{fig:avC1}       
\end{figure}
We now turn to spin correlations and
focus on their behavior in infinite-randomness phases, namely in the RS phase and at the Haldane-RS critical point, 
where the correlations between a typical pair of spins decay exponentially with the distance while the average correlations 
fall off algebraically.

First we examine  the distribution of end-to-end correlations, which consider
correlations between two end spins in an open chain with free boundary
conditions.  We rescale the extremely broad distributions at the critical point
$\delta_c=1$ and in the RS phase $\delta>1$ according to $-\ln C_1(L) \sim
L^\psi$, i.e. using the scaled variable $c=-\ln C_1(L)/L^\psi$ with SDRG theoretical values $\psi=\psi_c=1/3$
and $\psi=1/2$, respectively. The scaling plot
in Fig.~\ref{fig:C1_P} shows that a good data collapse for both $\delta=1.5$
and $\delta=1$ are achieved; 
furthermore, the data-collapsed distributions
for these two different universality classes 
are in a pretty similar shape.

There are no known analytical functions for the data-collapsed distributions
of the end-to-end correlations
in the RS phase and at the critical point that we consider here.
In Fig.~\ref{fig:C1_P}(b) and (c) we fit the collapsed distributions
to the form 
\be
    P(c)=Ac\exp(-Bc^2-Dc)\,,
    \label{eq:P_c}
\ee
where $A,B$ and $D$ are fitting parameters. 
This form  with $A=1/2$, $B=1/4$ and $D=0$ 
corresponds to the analytical result predicted for the infinite-randomness
critical point of the random transverse-field Ising spin chain~\cite{Fisher_Young}.
Here we include a finite linear term $-Dc$ in the exponential function
to achieve good fittings.

In an infinite-randomness phase, the average correlations $\overline{C}_1(L)$
are dominated by the rare event of the two end spins being strongly correlated.
Our data for the average end-to-end correlations at the critical point
$\delta_c=1$ and in the deep RS phase (with $\delta=1.4$ and $\delta=1.5$) are shown in
Fig.~\ref{fig:avC1} as a log-log plot.  For the critical point, we obtain
$\eta_1\approx0.7$, close to previous numerical result: $\eta_1=0.69$ found in
Ref.~\cite{DMRG1}.  The slope for  $\delta=1.5$ and for $\delta=1.4$ in the regime of large $L$ 
is about $\eta_1=1.1$, close to the analytical result $\overline{C}_1(L)\sim 1/L$ predicted for the
infinite-randomness critical point of the random transverse-field Ising spin
chain~\cite{Fisher_Young}.  We note that the linear dependence of $P(c)$ in
Eq.~(\ref{eq:P_c}) for small $c=-\ln C_1(L)/L^\psi$ (i.e. $P(c)\to Ac$ for $c\to 0$) is crucial for
obtaining the average end-to-end correlations $\overline{C}_1(L)\sim 1/L$ for the Ising case with $\psi=1/2$, as discussed in
Ref.~\cite{Fisher_Young}. 
However, we cannot ensure that 
Eq.~(\ref{eq:P_c}) is a correct function 
for the cases that we consider here.

\begin{figure}
\centering
\resizebox{0.8\columnwidth}{!}{\includegraphics{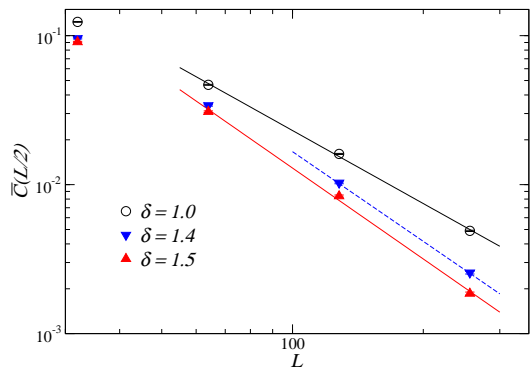}}
\caption{Average spin correlations $\overline{C}(r)$ at the longest distance $r=L/2$ 
at the critical point with $\delta=1.0$ and in the RS phase with with $\delta=1.5$ 
for several different chain lengths $L$. The solid lines are the best fits for data points from $L= 64$ to $L=256$, 
corresponding to
$\overline{C}\propto 1/L^\eta$ with $\eta=2.03$ for $\delta=1.5$ and  $\eta=1.62$ for $\delta=1$.
The dashed line corresponds to $\overline{C}\propto 1/L^2$ and fits well to the data
for $\delta=1.4$ in the regime of large $L$, too.}
\label{fig:avC}       
\end{figure}

Finally, we consider the bulk spin correlations.
In order to eliminate boundary effects and reduce finite-size effects, we here consider 
spin correlations at the largest distance $r=L/2$ in closed chains with PBC. 
In addition, different reference locations in the closed chain were sampled for the disorder average.
In Fig.~\ref{fig:avC}, our tSDRG results for the average bulk correlation in the RS phase
($\delta=1.4$ and $\delta = 1.5$) graphed versus the chain
length $L$ show a good agreement with the theoretical prediction: $\overline{C}(L)\sim 1/L^2$.
The average spin correlation function at $\delta_c=1$
in Fig.~\ref{fig:avC} shows an algebraic decay with
$\eta\approx 1.62$, which differs from
the inverse-square law in the RS phase.

The critical exponents for spin correlations found by our calculations are also summarized
in Table~\ref{Tab:exponent}.

\section{Summary and discussion}
\label{sec:discussion}
Using the tSDRG algorithm we have reproduced the zero-temperature phase diagram
of the spin-1 random Heisenberg chain depending on the randomness strength. 
We were able to obtain critical exponents in good agreement with the theoretical values,
both at the critical point and in the RS phase.
In comparison to previous DMRG results~\cite{DMRG1,DMRG2}, our tSDRG results show an overall better 
agreement with the theoretical predictions.
The advantage of the tSDRG approach lies in the straightforward implementation
for systems with periodic boundary conditions, which reduces finite size errors in bulk quantities.  
Furthermore, by comparing the tSDRG results for the mean spin correlations of the spin-1/2 random chain 
with the data obtained by non-approximate QMC calculations~\cite{1DJQ},  we have found that the tSDRG algorithm
can not only provide correct scaling forms but also achieve accurate numerical results. 
However, previous QMC simulations have uncovered logarithmic corrections to the
asymptotic $r^{-2}$ decay of the mean spin correlation in the spin-1/2 RS phase,
which are not captured by the tSDRG method (and is also not present in the SDRG analytical solution).
There have been attempts to further improve the accuracy of the tSDRG approach.
In Ref.~\cite{Chatelain} selections of blocks for the renormalization were adjusted to 
the specific models under consideration;
in Ref.~\cite{dMERA} optimization using variational energy minimization after coarse-graining
was introduced, as an extension of tSDRG and the multiscale entanglement renormalization ansatz (MERA)~\cite{MERA1,MERA2}.
An interesting question is whether these improved tSDRG methods can obtain the logarithmic corrections
found in QMC calculations.

\begin{acknowledgements}
We would like to thank Y.-J.~Kao, Y.-P.~Lin, J.-K.~Fang for previous collaboration.   
This work was supported by the
Ministry of Science and Technology (MOST) of Taiwan under Grants No.~107-2112-M-007-018-MY3,
No.~107-2112-M-004-001-MY3 and No.~108-2112-M-002-020-MY3. We also
acknowledge support from the NCTS.
\end{acknowledgements}

\section*{Author contribution statement}
P.~Chen and Y.~C.~Lin conceived and supervised the study.
Z.~L.~Tsai performed all the numerical calculations 
for spin-1 chains; Y.~C.~Lin performed the tSDRG calculations
for the spin-1/2 chain.  
All authors participated in the analyses of the results. 
Y.~C.~Lin wrote the paper with input from all authors.

\appendix
\section{Convergence test}\label{app}
\begin{table}[h]
\caption{\label{tab:chi} Average spin correlations $\overline{C}(L/2)$ with $L=64$ for $S=1/2$ and $S=1$ spin chains versus $\chi$.}
\begin{tabular}{lll}
\hline\noalign{\smallskip}
$\chi$ & $S=1/2$ & $S=1$ \\
\noalign{\smallskip}\hline\noalign{\smallskip}
5  & $0.00148(4)$  & $0.01229(14)$ \\
10 & $0.00282(5)$  & $0.02995(23)$  \\
20 & $0.00304(4)$  & $0.04859(24)$  \\
30 & $0.00299(4)$  & $0.04860(24)$  \\
40 & $0.00305(4)$  & $0.04864(24)$  \\
\noalign{\smallskip}\hline
\end{tabular}
\end{table}
Here in Table~\ref{tab:chi} we list the average spin correlations $\overline{C}(r)$ for spin $S=1/2$ and spin $S=1$ 
at $r=L/2$ in a closed chain of length $L=64$ with PBC and $\delta=1$,
against different values of bond dimensions $\chi$. 
At least 2000 disorder realizations and different reference locations in the closed chain were sampled
to get the average. 
The data converge already at $\chi=20$, both for $S=1/2$ and $S=1$.

\end{document}